\documentstyle[epsfig,rotating]{elsart}
\begin{document}
\begin{frontmatter}
\title{Coherent $\pi^o$-photoproduction from atomic nuclei}

\author[Basel]{B.~Krusche\thanksref{email}},
\author[Mainz]{J.~Ahrens},
\author[Mainz]{R.~Beck},
\author[Dubna]{S.~Kamalov},
\author[Giessen]{V.~Metag},
\author[Glasgow]{R.O.~Owens}
 and
\author[Juelich]{H.~Str\"oher}
\address[Basel]{Department of Physics and Astronomy, University of Basel,
Ch-4056 Basel, Switzerland}
\address[Mainz]{Institut f\"ur Kernphysik, Johannes-Gutenberg-Universit\"at 
                Mainz, D-55099 Mainz, Germany}
\address[Dubna]{Laboratory of Theoretical Physics, JINR Dubna, Russia}
\address[Giessen]{II. Physikalisches Institut, Universit\"at Giessen,
                  D-35392 Giessen, Germany}		
\address[Glasgow]{Department of Physics and Astronomy, University of Glasgow,
                  Glasgow G128QQ,U.K.}
\thanks[email]{corresponding author: Bernd.Krusche@unibas.ch}

\address[Juelich]{Institut f\"ur Kernphysik, Forschungszentrum J\"ulich GmbH,
52425 J\"ulich, Germany}

\begin{abstract}
Differential and total cross sections for coherent $\pi^o$-photo\-production 
from $^{12}$C, $^{40}$Ca, $^{93}$Nb and Pb targets have been measured
throughout the region of the $\Delta$(1232)-resonance. The experiments
were performed with the TAPS-detector at the Mainz accelerator MAMI.
The characteristic proportionality of the cross section to the square of the 
atomic mass number and to the nuclear mass form factor is clearly demonstrated.
The data allow for the first time detailed tests of model predictions for this
reaction. 
The comparison of the data to model predictions shows that the 
$\Delta$-nucleus interaction saturates: it is described for heavy nuclei with
the same potential parameters as for $^4$He.
\end{abstract}
\end{frontmatter}

\section{Introduction}
The study of meson photoproduction from atomic nuclei is mainly motivated by 
two strongly interconnected aspects namely possible medium modifications of the
excited states of the nucleon and the meson-nucleus interaction.
Three different reaction mechanisms can contribute for neutral mesons:
coherent production where the reaction amplitudes from all nucleons add
coherently and the nucleus remains in its ground state, incoherent meson
production with an excited nucleus in the final state, and breakup reactions
where at least one nucleon is knocked out of the nucleus. The theoretical
treatment of coherent meson production involves many fewer assumptions and
approximations than are needed for the description of the complicated final
states from breakup reactions. For the latter so far only semiclassical
calculations e.g. in the framework of transport models \cite{Lehr} or mean free
path Monte Carlo calculations \cite{Carrasco_MC} are available.
In light nuclei with only a few excited states, incoherent excitations of the
nucleus can be exploited as spin - isospin filters, but their treatment 
becomes very complicated for heavy nuclei with a high level density.

As long as final state interaction (FSI) of the mesons is neglected, the cross
section of breakup reactions scales with the mass number of the nuclei but the
differential cross section of the coherent process is proportional to the 
square of the mass number and the nuclear mass form factor. Due to the form factor
dependence, the coherent process is negligible compared to breakup reactions at
large momentum transfers. For this reason the in-medium properties of
higher lying nucleon resonances were only investigated via quasifree pion or eta
photoproduction (see e.g. \cite{Roebig,Krusche_2}). However, the situation is
different for the excitation of the lowest lying nucleon resonance, the
so-called $\Delta$-resonance ($P_{33}$(1232) in the usual notation \cite{PDG}).
The momenta transfered to nuclei in pion photoproduction to forward angles
are so small that the coherent process is dominant for heavy nuclei.
Furthermore, the elementary photoproduction of neutral pions from the nucleon
is well understood in this energy region and strongly dominated by the
excitation of the $\Delta$-resonance \cite{Drechsel2}.
  
Coherent $\pi^o$-photoproduction has been extensively treated in the framework 
of different models. One group of models, which mainly aim at the study of the 
pion-nucleus interaction and the in-medium properties of the pion, use the 
distorted wave impulse approximation (DWIA) 
(see e.g. \cite{Girijia,Kamalov,Boffi,Chumbalov}). 
Starting from the elementary pion-nucleon amplitude this approach dynamically 
takes into account final state interactions (FSI) of the pion in the nuclear 
medium but neglects medium modifications of the resonance properties (position 
and widths). 
On the other hand the in-medium properties of the $\Delta$-resonance have been
intensively studied in the framework of the $\Delta$-hole approach with special
attention to the $\Delta$ and pion dynamics but mostly without non-resonant 
contributions in the elementary production process 
(see e.g. \cite{Oset,Koch1,Korfgen}). In the case of $^{12}$C Takaki et al.
\cite{Takaki} have extended this calculations to incoherent contributions from
low-lying nuclear excitations. Since the $\Delta$-hole calculations are
numerically quite involved, they have been mostly restricted to light nuclei.
Carrasco et al. {\cite{Carrasco} have tried to overcome this difficulty with a
local approximation of the $\Delta$-hole model, which made calculations feasible
even for lead.  
Prompted by the growing interest in medium modifications of hadrons 
and the expected availability of new, precise data, several extensions
of the existing models have been discussed recently. In the framework of the
DWIA approach Drechsel and coworkers \cite{Drechsel} presented a calculation 
which starts from their Unitary Isobar Model for the elementary reaction
\cite{Drechsel2} and includes a phenomenological parametrization of the 
$\Delta$ self-energy.  Peters et al. \cite{Peters} developed a relativistic
non-local model which includes medium modifications in the production
operator of the $\Delta$-resonance. Abu-Raddad et al. \cite{Raddad} emphasized
the ambiguities arising already from the extrapolation of the elementary
amplitude off the mass shell in the relativistic impulse approximation.

On the experimental side progress was much slower. In a very early attempt
Schrack et al. \cite{Schrack} measured unnormalized pion yield curves for light 
to medium mass nuclei. Since then differential and total cross sections
have been measured for carbon and calcium nuclei in the threshold region
\cite{Gothe,Koch} but only one attempt was made to measure the cross section
throughout the $\Delta$-resonance for carbon \cite{Arends}.
Only recently measurements up to much higher incident photon energies were 
reported for deuterium and helium nuclei \cite{Krusche_1,Rambo,Bellini,Tieger}. 

\section{Experiment and analysis}
In this paper we present the results of measurements of coherent 
$\pi^o$-photo\-production from  $^{12}$C-, $^{40}$Ca-, $^{93}$Nb- and 
Pb-targets throughout the region of the $\Delta$-resonance. The experiments 
were carried out at the Glasgow tagged photon facility installed at the Mainz 
Microtron (MAMI) with the TAPS-detector \cite{Novotny,Gabler}. Details of the 
experimental setup and the data analysis are summarized in \cite{Krusche_1}.

The separation of the coherent and incoherent parts is only possible via their 
different reaction kinematics. 
This was done by a comparison of the cm energy $E_{\pi}(\gamma_1\gamma_2)$ 
of the pion derived from the measurement of energy and momentum of its decay 
photons to the energy $E_{\pi}(E_{\gamma})$
derived from the incident photon energy $E_{\gamma}$ under the assumption of 
coherent pion production \cite{Krusche_1}.
The situation for heavy nuclei differs in two aspects from the deuterium case
discussed in \cite{Krusche_1}. The larger nucleon binding energies and the
stronger effects of Fermi motion result in a much better separation of coherent
and breakup events in missing energy. On the other hand no model independent
simulation of the shape of the contribution from the complicated final states
of non-coherent events in missing energy is possible. Therefore only the shape
of the coherent contribution was simulated and compared to the data at positive
values of the pion energy difference, i.e. in the region where non-coherent 
components can only contribute due to finite resolution effects.
Contributions from breakup
reactions are almost completely removed in this way but incoherent
excitations to low-lying nuclear states are only incompletely suppressed.
Due to the shape of the angular distributions and the $A^2$-dependence of the
cross section of the coherent process, residual incoherent background is mainly
a concern for light nuclei and large pion angles. Examples of the measured
energy difference spectra
($\Delta E = E_{\pi}(\gamma_1\gamma_2)-E_{\pi}(E_{\gamma}))$  
and the Monte Carlo simulation are shown in fig. 
\ref{fig:1}. The measured shape is almost perfectly reproduced by the 
simulation at low incident photon energies and small pion angles. At higher 
energies or larger angles non-coherent contributions become visible.    
\begin{figure}[h]
\begin{center}
\epsfig{figure=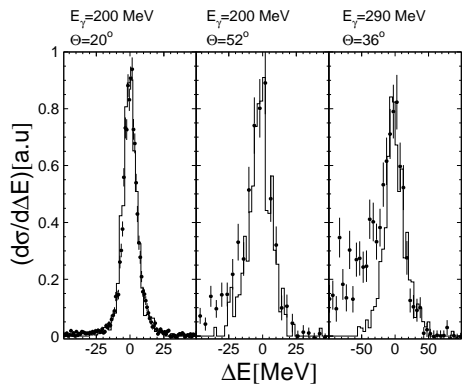,width=6.cm}\\[0.5cm]
\caption{Pion energy difference spectra for the reaction Pb$(\gamma ,\pi^o)$X.
Symbols with error bars represent the data, the histograms a Monte Carlo
simulation of the coherent reaction Pb$(\gamma ,\pi^o)$Pb. The energy and angle
combinations correspond to the first and second maximum of the 
angular distribution for the incident photon energy $E_{\gamma}=$200 MeV and 
to the second maximum at $E_{\gamma}=$290 MeV (compare fig. \protect\ref{fig:4}).
\label{fig:1}}
\end{center}
\end{figure}

The absolute normalization of the cross sections was done as in \cite{Krusche_1}.
The total overall uncertainty due to analysis cuts, target thickness and photon
flux amounts to 3\%. The systematic uncertainty of the simulation of the
detection efficiency is estimated to be 10\%.

\section{Results and discussion}
The quality of the data obtained is shown in fig. \ref{fig:2}, 
where the differential cross sections averaged over photon energies from 
200 - 290 MeV are plotted versus the momentum $q_A$ transfered to the nucleus.  

The coherent cross section for spin zero nuclei can be written in the most 
simple plane wave impulse approximation (PWIA) \cite{Drechsel} as:
\begin{equation}
\frac{d\sigma_{PWIA}}{d\Omega}(E_{\gamma},\Theta_{\pi}) =
\frac{s}{m_N^2}A^2
\frac{d\sigma_{NS}}{d\Omega}(E_{\gamma}^{\star},\Theta_{\pi}^{\star})
F^2(q_A)
\label{eq:1}
\end{equation}
\begin{equation}
\frac{d\sigma_{NS}}{d\Omega}(E_{\gamma}^{\star},\Theta_{\pi}^{\star}) =
\frac{1}{2}\frac{q^{\star}}{k^{\star}}
|{\cal{F}}_2(E_{\gamma}^{\star},\Theta_{\pi}^{\star})|^2 
sin^2(\Theta_{\pi}^{\star}),
\label{eq:3}
\end{equation}
where $E_{\gamma}$ and $\Theta_{\pi}$ are
the incident photon energy and the pion polar angle
in the {\em photon-nucleus} cm-system, $A$ is the atomic mass number, $m_N$ 
is the nucleon mass and $F(q_A)$ the nuclear mass form factor  The total energy 
$\sqrt{s}$ of the photon-nucleon 
pair, the photon energy and momentum $E_{\gamma}^{\star}$, $k^{\star}$, and the
pion angle and momentum $\Theta_{\pi}^{\star}$, $q^{\star}$ in the 
{\em photon-nucleon} cm-system are evaluated from the average momentum 
$\vec{p_N}$ 
of the nucleon in the factorization approximation $\vec{p}_N=\vec{q}_A(A-1)/2A$.
The spin-independent elementary cross section $d\sigma_{NS}/d\Omega$
is calculated from the standard Chew-Goldberger-Low-Nambu (CGLN) amplitude 
${\cal{F}}_2$ \cite{Chew} taken from \cite{Drechsel2} and averaged over 
proton und neutron numbers.

\begin{figure}[t]
\begin{center}
\epsfig{figure=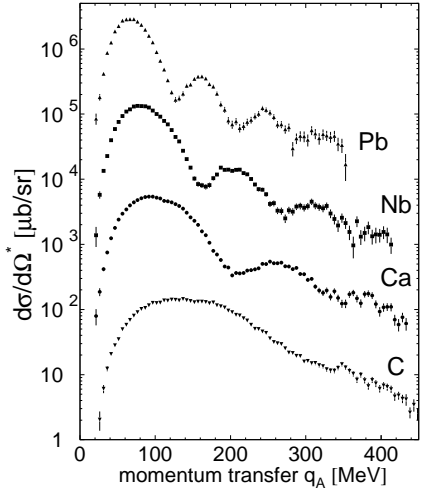,width=5.5cm}\\[0.5cm]
\caption{Differential cross sections for A$(\gamma ,\pi^o)$A
averaged over incident photon energies from 200 - 290 MeV as 
function of the momentum transfer.
The scale corresponds to the carbon data, the Ca-, Nb- and Pb-data are
scaled up by factors 10, 100, 1000, respectively.
\label{fig:2}}
\end{center}
\end{figure}
\begin{figure}[ht]
\begin{center}
\epsfig{figure=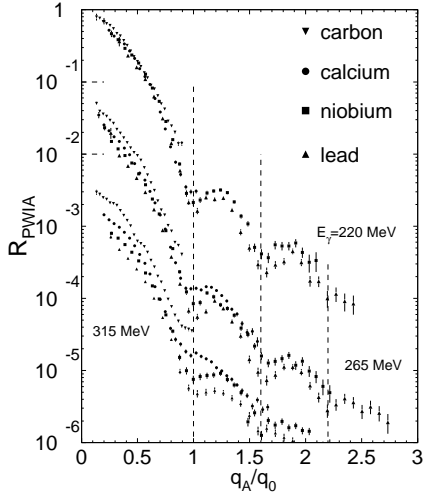,width=5.5cm}\\[0.5cm]
\caption{Ratio $R_{PWIA}$ (see eq. \ref{eq:4}) as function of $q_A/q_0$ where 
$q_0$ is the  momentum transfer at the first minimum of the form factor.
The scale corresponds to the 220 MeV data, the 265, 315 MeV data are scaled 
down by factors 10, 100. 
\label{fig:3}}
\end{center}
\end{figure}

The influence of the form factor and the $sin^2$-term, which forces the 
forward cross section to zero, is clearly visible in fig. \ref{fig:2}. 
The data allow also a direct check of the characteristic 
$A^2$-dependence. For this purpose we define the ratio: 
\begin{equation}
R_{PWIA}=
(\frac{d\sigma_{exp}}{d\Omega})/
[\frac{s}{m_N^2}A^2
(\frac{d\sigma_{NS}}{d\Omega})]=
F^2(q_A)(\frac{d\sigma_{exp}}{d\Omega})/
(\frac{d\sigma_{PWIA}}{d\Omega})
\label{eq:4}
\end{equation}
which is plotted in fig. \ref{fig:3} versus $q_A/q_0$, where $q_0$ 
is the  momentum transfer at the first diffraction minimum. 
As long as PWIA is valid, the ratios should equal the square
of the nuclear mass form factors, which are similar for all nuclei as 
function of $q_A/q_0$.
For the lowest incident photon energies around 220 MeV the ratios for all 
nuclei are indeed very similar and close to
unity for small momentum transfers which demonstrates the approximate validity 
of the PWIA in this region. For higher photon energies we expect 
significant FSI effects since the pions then have a much larger cross section 
for the excitation of the $\Delta$-resonance. The data show this effect as a
decrease of the cross section ratio with mass number.
 
A detailed investigation of the FSI effects and possible medium modifications of
the $\Delta$-resonance requires an analysis far beyond PWIA. As a first step 
the angular distributions for $^{12}$C, $^{40}$Ca and Pb are compared in 
fig. \ref{fig:4} to results from the model of Drechsel et al. \cite{Drechsel}.
The three curves in the figure correspond to PWIA, DWIA and the full model.
In addition to pion FSI this includes also the medium modification of the 
$\Delta$-resonance properties due to the $\Delta$-nucleus interaction via a 
phenomenological parametrization of the $\Delta$ self-energy.
The $\Delta$ self-energy was fitted to the $^4$He$(\gamma, \pi^o)^4$He reaction 
\cite{Drechsel,Rambo} and this parametrization was used without modification to
calculate the cross sections for C, Ca and Pb.

\begin{figure}[ht]
\begin{center}
\epsfig{figure=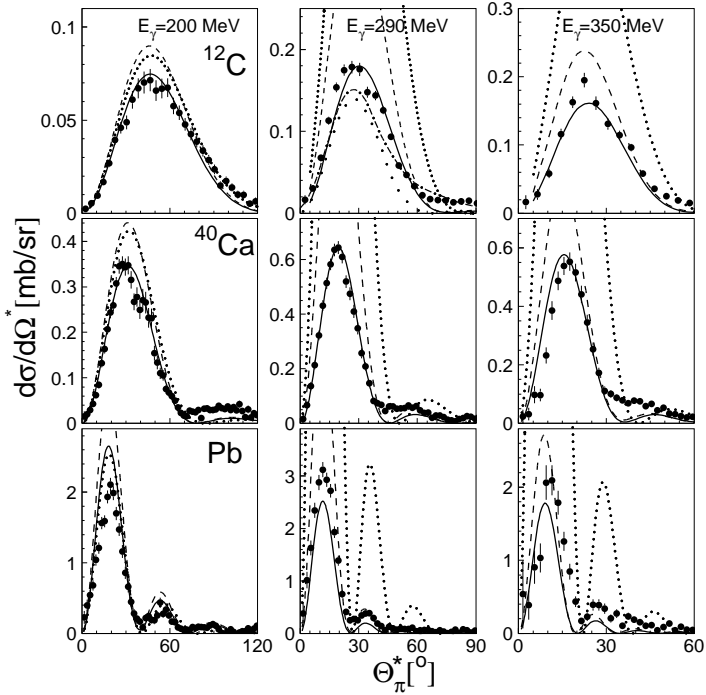,width=9.3cm}\\[0.5cm]
\caption{Differential cross sections for $^{12}$C$(\gamma ,\pi^o)^{12}$C,
$^{40}$Ca$(\gamma ,\pi^o)^{40}$Ca, and Pb$(\gamma ,\pi^o)$Pb compared
to the predictions from Drechsel et al. 
\protect\cite{Drechsel}. 
Dotted lines:
PWIA, dashed lines: DWIA, full lines: DWIA with $\Delta$-self energy fitted to
$^4$He cross sections. For the carbon data at 290 MeV the predictions from
ref. \protect\cite{Takaki} for the coherent reaction (wide space dotted) and 
coherent plus incoherent excitation of low lying states (dash-dotted) are also
shown.
\label{fig:4}}
\end{center}
\end{figure}

At low incident photon energies ($E_{\gamma}=200$ MeV) the difference between 
PWIA, DWIA and additional $\Delta$-modification is small. However, the 
cross sections are strongly overestimated around the $\Delta$-resonance 
position by the PWIA and DWIA calculations. This in contrast to our results
for coherent photoproduction from the deuteron \cite{Krusche_1} which are 
in good agreement with the DWIA calculation. Reasonable agreement for the
heavier nuclei is only 
achieved when the $\Delta$-nucleus interaction is taken into account. 
The $\Delta$-self energy extracted from the $^4$He data for this incident 
photon energy ($E_{\gamma}=290$ MeV) is $Re(V)\approx 19$ MeV and 
$Im(V)\approx -33$ MeV \cite{Drechsel}, corresponding to a significant effective
broadening of the resonance by 66 MeV.
Based on a comparison of their prediction to the few data then available 
for $^{12}$C$(\gamma ,\pi^o )^{12}$C it was suggested by Drechsel et al. 
\cite{Drechsel} that the $\Delta$-nucleus interaction already saturates for 
$^4$He. The present data demonstrate that indeed the A-dependence of the 
potential is not large since the agreement between model predictions and data
is very good for carbon and calcium and still reasonable for lead. 

\begin{figure}[ht]
\begin{center}
\epsfig{figure=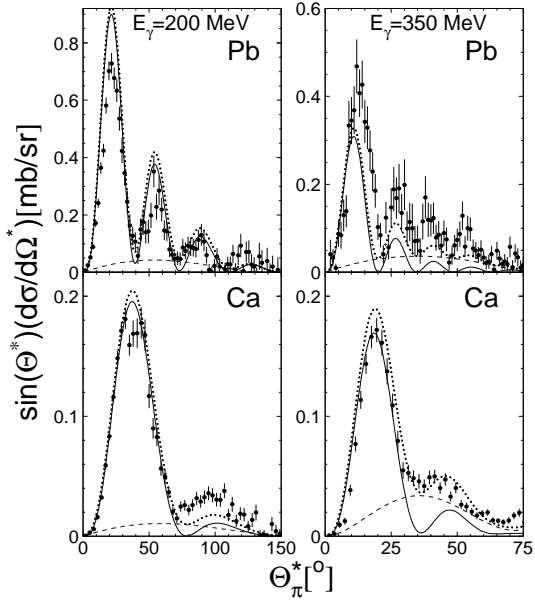,width=7.cm}\\[0.5cm]
\caption{Differential cross sections for
$^{40}$Ca$(\gamma ,\pi^o)^{40}$Ca, and Pb$(\gamma ,\pi^o)$Pb multiplied
by $sin(\theta^{\star})$. The full curves are the predictions from
Drechsel et al.\protect\cite{Drechsel} for the coherent process. The dashed 
lines indicate the estimate (see text) for possible incoherent contaminations 
of the data. The dotted lines show the sum of the prediction for the coherent 
cross section and the incoherent background.
\label{fig:5}}
\end{center}
\end{figure}
A more detailed comparison of the data to model predictions requires some
remarks concerning possible background from incoherent excitations of
nuclear levels. It was pointed out by Takaki et al. \cite{Takaki} that such
components will have a significant effect on the total cross section
since they peak at more central pion angles which gives them a large weight for 
the total cross section. The prediction of the angular distribution for $^{12}$C
at $E_{\gamma}$=290 MeV from their $\Delta$-hole model with and without 
incoherent excitations is also compared to the data in fig. \ref{fig:4}. 
The model underestimates the data, even when incoherent excitations are 
included. Some examples of the differential cross sections weighted with 
$sin(\Theta_{\pi}^{\star})$ are compared to the predictions from \cite{Drechsel}
in fig. \ref{fig:5}. This presentation shows directly the contribution of 
different angular ranges to the total cross section.
\begin{figure}[ht]
\begin{center}
\epsfig{figure=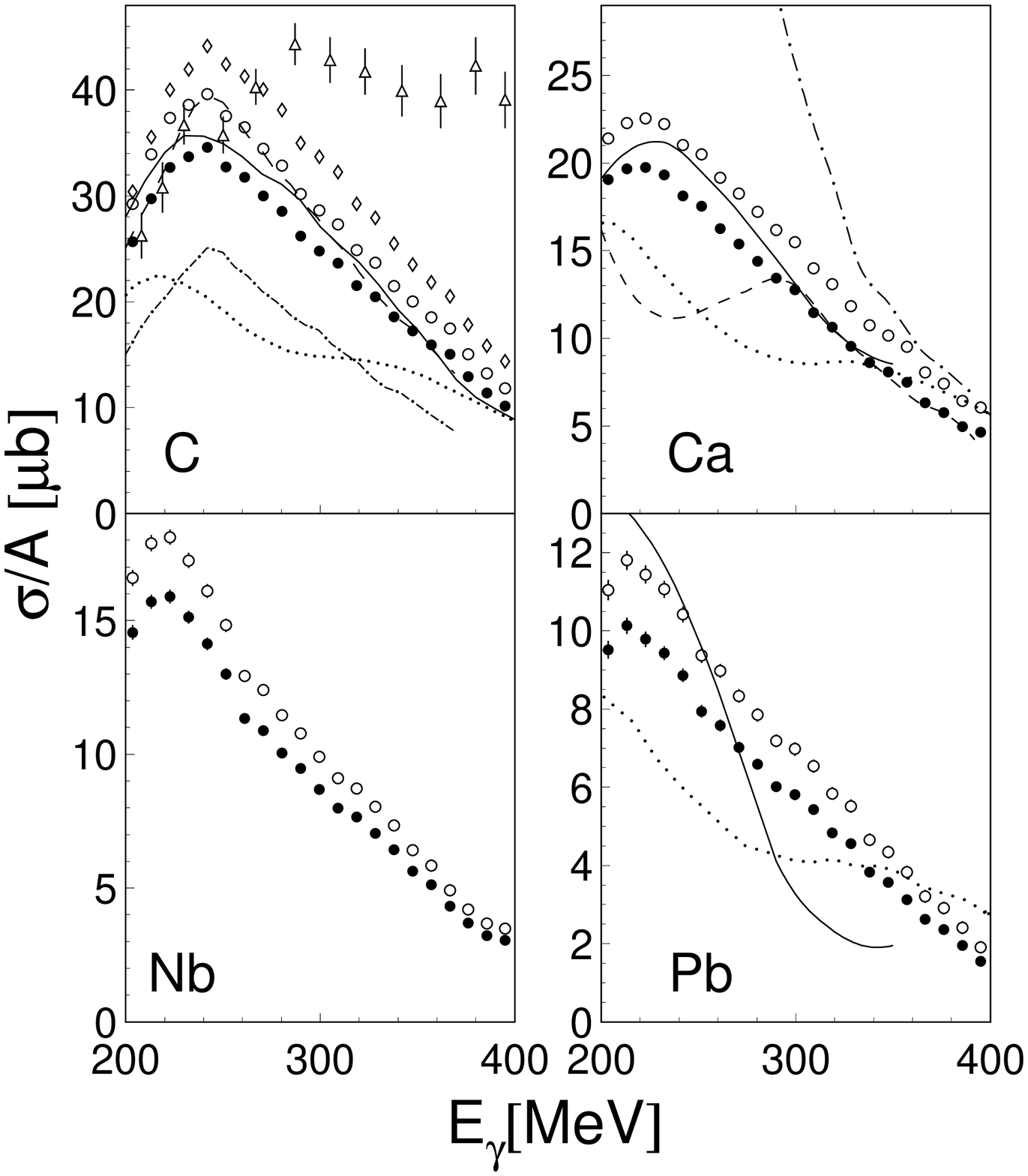,width=10.cm}\\[0.5cm]
\caption{Total cross sections for $^{12}$C$(\gamma ,\pi^o)^{12}$C,
$^{40}$Ca$(\gamma ,\pi^o)^{40}$Ca, and Pb$(\gamma ,\pi^o)$Pb. Open circles:
cross section obtained by integration of the differential cross sections up to
momentum transfers of  
330 MeV (carbon), 340 MeV (calcium) and 360 MeV (niobium, lead). Filled circles:
with correction for incoherent contamination (see text).
Full curves: prediction by Drechsel et al. \protect\cite{Drechsel}. The dotted
lines are predictions from Carrasco et al. \protect\cite{Carrasco}.
For carbon the open diamonds show the total cross section by integration over
the full angular range, the open triangles are data from ref. 
\protect\cite{Arends}. The
dash-dotted and dashed curves are the prediction from \protect\cite{Takaki}
for the coherent and coherent plus incoherent processes, respectively.
For calcium the dashed and dash-dotted lines are the predictions from ref. 
\cite{Raddad} for vector (dashed) and tensor (dash-dotted) parametrization of 
the elementary amplitude. 
\label{fig:6}}
\end{center}
\end{figure}
A first estimate of the total 
coherent cross sections has been obtained by integration of the angular 
distributions up to momentum transfers of 330, 340, 360, 360 MeV for 
$^{12}$C, $^{40}$Ca, $^{93}$Nb and Pb, respectively. These momentum transfers
(see fig. \ref{fig:2}) correspond to the position of the first (C), second (Ca), 
third (Nb) and fourth (Pb) minimum of the form factor. The results of the
integration are shown in fig. \ref{fig:6}.
This restriction of the angular ranges is reasonable since the model
\cite{Drechsel} predicts that approximately 98\% of the total coherent cross
section is contained in these ranges, but it significantly suppresses incoherent
background which occurs at higher momentum transfers (as shown for 
$^{12}$C in fig. \ref{fig:6}).
Since the total cross sections obtained in this way (open circles in fig. 
\ref{fig:6}) may still include residual
background from incoherent processes in the chosen angular ranges they represent
only an {\it upper} limit of the coherent cross section.
An estimate of these incoherent contributions has been obtained under the 
assumption that the differential cross section in the diffraction minima of the
coherent process is entirely due to incoherent processes and that these 
contributions can be smoothly interpolated (see fig \ref{fig:5}). The estimated
corrections are in the range 12\% - 16\% at $E_{\gamma}$=200 MeV and 14\% - 26\%
at $E_{\gamma}$=350 MeV. These corrections are certainly overestimates since the
DWIA calculations from \cite{Drechsel} indicate that the coherent cross section
is not zero at the diffraction minima and this will be further accentuated by
the experimental resolution. The corrected data (filled circles in fig.
\ref{fig:6}) thus represent a {\it lower} limit of the coherent cross section. 
The results with this correction are compared to previous
data and model predictions in fig. \ref{fig:6}. The only data available in this
energy region so far are from the experiment of Arends et al. \cite{Arends} 
who used an active target for the separation of coherent and non-coherent
events in $\pi^o$-photoproduction from $^{12}$C. However, in that experiment
events from breakup reactions e.g. with a low energy neutron in the final state 
could not be completely suppressed. The comparison to the new data shows that   
at higher energies such contaminations were obviously large. The model
predictions by Drechsel et al. \cite{Drechsel} are in good agreement with
the carbon and calcium data but less so for lead. The local approximation to
the $\Delta$-hole model by Carrasco et al. \cite{Carrasco} underestimates all
total cross sections and shows a different energy dependence. The $\Delta$-hole
calculation for $^{12}$C by Takaki et al. \cite{Takaki} also 
underestimates the data on an absolute scale, although the energy dependence is
quite well reproduced. Note that in this case the data integrated over the 
full angular range (open diamonds in fig. \ref{fig:6}) represent a lower
limit for the cross section of coherent plus incoherent events from 
\cite{Takaki} since part of the incoherent contribution in the data is 
suppressed by the missing energy cut. However, the prediction underestimates 
even this lower limit. Finally the predictions
from \cite{Raddad} for $^{40}$Ca with vector and tensor parametrization of the
elementary amplitude both disagree with the data. 

\section{Conclusion}
Coherent $\pi^o$-photoproduction from nuclei was studied in detail throughout
the $\Delta$-resonance region. The characteristic features of the coherent
process, the proportionality of the cross section to $sin^2(\Theta_{\pi})$, 
to the nuclear mass form
factor and to the square of the nuclear mass numbers are demonstrated.
The data are quite well reproduced by simple PWIA calculations at incident 
photon energies around 200 MeV, but at higher photon energies 
distortion effects become large.
Predictions for the coherent cross section are available from many detailed
theoretical studies, however the results from different models are not at all in
agreement (see fig. \ref{fig:6}). Best agreement with the data is found for the
model by Drechsel et al. \cite{Drechsel}.
A comparison of the experimental results to these calculations demonstrates 
their 
sensitivity to medium modifications of the $\Delta$-isobar. It is shown, that 
reasonable agreement between data and model predictions for all nuclei is 
achieved with a parametrization of the $\Delta$-self energy fitted to the 
differential cross sections of the reaction $^4$He$(\gamma ,\pi^o)^4$He. This is
an indication that the $\Delta$-nucleus potential is not strongly
dependent on the nuclear mass number but saturates already for $^{4}$He.

{\bf Acknowledgments}
We wish to acknowledge the outstanding support of the accelerator group of
MAMI, as well as many other scientists and technicians of the Institute
f\"ur Kernphysik at the University of Mainz. 
We thank M. R\"obig-Landau for his contribution to the calibration and analysis
of the data. 
We gratefully acknowledge many stimulating discussions with L. Tiator.
This work was supported by Deutsche Forschungsgemeinschaft (SFB 201)
and the UK Science and Engineering Research Council.

\end{document}